# Non-reciprocity in elastic unpolarized neutron scattering by systems with non-coplanar magnetic structure

D.A. Tatarsky, O.G. Udalov, A.A. Fraerman

*Institute for physics of microstructures RAS, GSP-105, Nizhny Novgorod 603950, Russia*

The paper provides a theoretical analysis for elastic unpolarized neutron scattering by systems having a non-coplanar magnetic induction spatial distribution. It is shown that reciprocity does not hold for neutron scattered by these systems. The scattering of unpolarized neutrons by a system of three magnetic mirrors and by a ferromagnetic particle in a "vortex" state is also explored.

## I. INTRODUCTION

It is well known that time reversal symmetry of the particle motion equations leads to the following equality for the elastic scattering cross section $\sigma(\vec{k},\vec{k}',\vec{B})$ [1]

$$\sigma(\vec{k},\vec{k}',\vec{B}) = \sigma(-\vec{k}',-\vec{k},-\vec{B}), \qquad (1)$$

where $\vec{k}$ and $\vec{k}'$ are wave vectors of the incident and scattered particles, and $\vec{B}$ is the magnetic field in the system. Formula (1) is widely mentioned in different publications. Following [1,2] in this article equation (1) is referred to as the "reciprocity theorem". Often the adjectives "reciprocal" and "time-reversal invariant" are used to describe scattering processes that obey the equation $\sigma(\vec{k},\vec{k}') = \sigma(-\vec{k}',-\vec{k})$ [3-6]. The "reciprocity theorem" shows that the elastic scattering is always reciprocal in the absence of the "external" magnetic field in the system. Non-reciprocal elastic scattering can appear only in the system with the magnetic field (or spontaneous magnetic moment in the system) [3]. In this case the following holds

$$\sigma(\vec{k},\vec{k}',\vec{B}) \neq \sigma(-\vec{k}',-\vec{k},\vec{B}). \qquad (2)$$

Forward and time-reversed scattering processes are depicted in fig. 1.

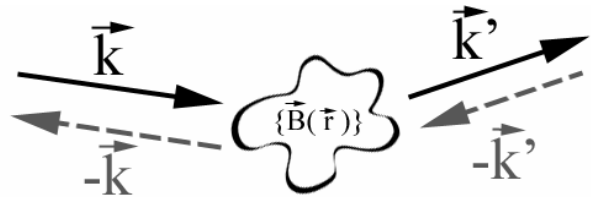

FIG. 1. Forward and time-reversed scattering processes.

Non-reciprocal scattering is well known in optics, for example, the non-reciprocal Faraday element [7]. As another example, the reflection of electromagnetic waves by antiferromagnets in the presence of an external magnetic field follows the inequality $R(\alpha) \neq R(-\alpha)$, where $R$ is the reflection coefficient and $\alpha$ is the incident angle [5,8-10].

According the microscopic approach, the reason for non-reciprocal scattering mentioned above is relativistic interactions (Lorentz force and spin-orbit coupling) [2,7]. Interaction between a neutron and a magnetic field has essentially different nature and is described by the Pauli term $-(\hat{\vec{\mu}} \cdot \vec{B})$ ($\hat{\vec{\mu}}$ is neutron magnetic moment) [11]. In contrast to Lorentz force and spin-orbit, magneto-dipole interaction has an additional symmetry. Rotation of magnetic field at each point of the space by the same angle (see the next section for details) can be compensated by the spin coordinate system rotation [12,13]. Therefore in case of unpolarized neutrons (for which spin coordinate turning does not lead to any physical changes) the scattering is independent of magnetic field rotation. Such additional symmetry results in always reciprocal scattering of unpolarized neutron by the systems with collinear magnetic field distribution in contrast to optics. In the paper we show that non-reciprocal elastic scattering of unpolarized neutron appears only in the systems with non-coplanar magnetic field spatial distribution and only beyond the first Born approximation.

The peculiar properties of the systems with non-coplanar magnetic field distribution are discussed in the corresponding work [14]. In this work specific geometry of scattering and specific magnetic structure is considered, namely the reflection of polarized neutrons by the media with ferromagnet spiral. It is theoretically demonstrated that an angular dependence of specular neutron reflection from semi infinite media with a cone spiral magnetic structure has an additional peculiarity in comparison with an angular dependence of reflection from the media with coplanar magnetic spiral. In the work [15] transmission of neutrons through the multilayer structure with non-coplanar magnetic spatial distribution is explored. The work shows that non-reciprocal transmission in one dimensional structure appears due to neutron absorption. Realization of the non-coplanar magnetic induction spatial distribution in multilayer system is difficult due to demagnetizing field existence. This problem is discussed in the paper [15]. In the present work a more general situation is considered, touching upon not only the specular reflection or transmission, but also the diffusive scattering. The problem of non-reciprocal elastic neutron scattering is considered in this paper in terms of symmetry. Also scattering by the two experimentally feasible systems is explored, namely, a nanoparticle with vortex magnetization distribution and a system of three magnetic mirrors.

The paper is organized as follows. First, the non-reciprocity of elastic neutron scattering is explored by means of phenomenological approach. Next, we analyze neutron elastic scattering by ferromagnetic nanoparticles with a "vortex" magnetization distribution. Finally, we explore the problem of neutron reflection by the ensemble of three magnetic mirrors.

## II. PHENOMENOLOGICAL CONSIDERATION

Consider unpolarized neutron elastic scattering by magnetic heterogeneities. The total cross section is denoted as $\sigma(\vec{k},\vec{k}',\{B(\vec{r})\})$, and is related to the scattering amplitude $f(\vec{k},\vec{k}',\{B(\vec{r})\})$ by

$$\sigma(\vec{k},\vec{k}',\{\vec{B}(\vec{r})\}) = Tr(f^+(\vec{k},\vec{k}',\{\vec{B}(\vec{r})\})$$
$$\times f(\vec{k},\vec{k}',\{\vec{B}(\vec{r})\}))/2. \qquad (3)$$

Here, $Tr$ refers to the trace over spin indexes. To get some general properties of $\sigma(\vec{k},\vec{k}',\{B(\vec{r})\})$ let us analyze first the symmetry of neutron interaction with magnetic field in the media. As it was mentioned in the introduction this interaction is described by the term $U_{MD} = -(\hat{\vec{\mu}} \cdot \vec{B})$ [12,13] and depends only on the mutual angle between neutron spin and magnetic induction. Denote by $\hat{R}_{\vec{n}}^{\alpha}$ the operator of vector rotation by the angle $\alpha$ around $\vec{n}$-vector. $\hat{R}_{\vec{n}}^{\alpha}$ does the following transformation of the magnetic field $\vec{B}(\vec{r})$

$$\hat{R}_{\vec{n}}^{\alpha}\vec{B}(\vec{r}) = B_x(\vec{r})\hat{R}_{\vec{n}}^{\alpha}\vec{x}_0 + B_y(\vec{r})\hat{R}_{\vec{n}}^{\alpha}\vec{y}_0 + B_z(\vec{r})\hat{R}_{\vec{n}}^{\alpha}\vec{z}_0 . \qquad (4)$$

Here $\vec{x}_0, \vec{y}_0, \vec{z}_0$ are unit vectors in $x,y,z$ directions, respectively, $\vec{r}$ is a position vector, $B_{x,y,z}(\vec{r})$ represents magnetic induction projections on $x,y,z$ axes. Such a transformation turns magnetic induction at each point of space by the same angle $\alpha$ around $\vec{n}$ vector, but does not move the origin of $\vec{B}$ vectors (does not transform $\vec{r}$). Denote by $\hat{S}_{\vec{n}}^{\alpha}$ the operator of spin coordinate system rotation by the angle $\alpha$ around $\vec{n}$ vector. $\hat{S}_{\vec{n}}^{\alpha} = \cos(\alpha/2) + i(\vec{n}\hat{\vec{\sigma}})\sin(\alpha/2)$ where $\hat{\vec{\sigma}}$ – Pauli matrix vector. It easy to show that effect of $\hat{R}_{\vec{n}}^{\alpha}$ transformation on $U_{MD}$ can be compensated by $\hat{S}_{\vec{n}}^{-\alpha}$ transformation.

$$(\hat{S}_{\vec{n}}^{-\alpha})(\hat{\vec{\mu}} \cdot \hat{R}_{\vec{n}}^{\alpha}\vec{B})(\hat{S}_{\vec{n}}^{-\alpha})^{-1} = (\hat{\vec{\mu}} \cdot \vec{B}) \tag{5}$$

Since we consider unpolarized neutron beams, the rotation of spin coordinates does not lead to any physical changes in the system. Therefore as it follows from (5) $\hat{R}_{\vec{n}}^{\alpha}$ also does not change any measurable values and consequently $\sigma(\vec{k},\vec{k}',\{B(\vec{r})\})$ should not depend on rotation of magnetic induction.

$$\sigma(\vec{k},\vec{k}',\{\hat{R}_{\vec{n}}^{\alpha}\vec{B}(\vec{r})\}) = \sigma(\vec{k},\vec{k}',\{\vec{B}(\vec{r})\}) \tag{6}$$

According to the equation, non-reciprocity can not appear in the system with collinear or non-collinear but coplanar magnetic induction spatial distribution.

$$\sigma(\vec{k},\vec{k}',\vec{B}) = \sigma(-\vec{k}',-\vec{k},-\vec{B})$$
$$= \sigma(-\vec{k}',-\vec{k},\hat{R}_{\vec{g}}^{\pi}(-\vec{B})) = \sigma(-\vec{k}',-\vec{k},\vec{B}) \tag{7}$$

where $\vec{g}$ is directed perpendicular to the plane in which the magnetic field vector lays.

Further assume that the scattering is weak enough to be expanded in a series over the magnetic induction. Requirement of invariance with respect to rotation of $\vec{B}$ strongly restricts possible form of neutron scattering cross section. Particularly

$$\sigma(\vec{k},\vec{k}',\{\vec{B}(\vec{r})\}) = \sigma_0(\vec{k},\vec{k}')$$
$$+ \int Q_1(\vec{k},\vec{k}';\vec{r}_1,\vec{r}_2)(\vec{B}_1 \cdot \vec{B}_2)d\vec{r}_1 d\vec{r}_2$$
$$+ \int Q_2(\vec{k},\vec{k}';\vec{r}_1,\vec{r}_2,\vec{r}_3)(\vec{B}_1 \cdot [\vec{B}_2 \times \vec{B}_3])d\vec{r}_1 d\vec{r}_2 d\vec{r}_3 + ... \tag{8}$$

where $Q_{1,2}$ are scalar functions, $\vec{B}_j \equiv \vec{B}(\vec{r}_j)$. $\sigma_0(\vec{k},\vec{k}')$ is a part of cross section independent of the magnetic field. Non-reciprocity of the scattering, as follows from Eq. (1) and Eq. (8), can be expressed as

$$\sigma(\vec{k},\vec{k}',\{\vec{B}(\vec{r})\}) - \sigma(-\vec{k}',-\vec{k},\{\vec{B}(\vec{r})\})$$
$$= 2\int Q_2(\vec{k},\vec{k}';\vec{r}_1,\vec{r}_2,\vec{r}_3)(\vec{B}_1[\vec{B}_2 \times \vec{B}_3])d\vec{r}_1 d\vec{r}_2 d\vec{r}_3 + ..., \tag{9}$$

It can be seen from Eq. (9) that non-reciprocity of unpolarized neutron elastic scattering exists only in a system with a non-coplanar magnetic field spatial distribution, for which the scalar triple product $(\vec{B}(\vec{r})[\vec{B}(\vec{r}_1) \times \vec{B}(\vec{r}_2)])$ is not zero. Eq. (9) also shows that non-reciprocity is cubical in the magnetic field. This means that the phenomena can be found only beyond the first Born approximation.

III. ELASTIC NEUTRON SCATTERING BY FERROMAGNETIC NANOPARTICLE WITH "VORTEX" MAGNETIZATION DISTRIBUTION

One of the possible realizations of the system with non-coplanar magnetic field spatial distribution is a nanoparticle with vortex magnetization distribution. Within the frame of perturbation theory, consider neutron scattering by a cylindrical ferromagnetic particle with radius $R_0$ and width $h_0$. Under a certain relation of $R_0$ and $h_0$, the particle has a vortex magnetic structure [16]. In this case, the magnetization in the disc is determined by the expression

$$\vec{M}(\rho,\varphi) = \begin{pmatrix} -\sqrt{1-m_z^2(\rho)}\sin(\varphi+\varphi_0) \\ \sqrt{1-m_z^2(\rho)}\cos(\varphi+\varphi_0) \\ m_z(\rho) \end{pmatrix}. \tag{10}$$

Here, it is supposed that the z-axis is directed perpendicular to the base of the cylinder (Fig. 2a). $\varphi_0$ takes on values 0 and $\pi$, corresponding to a counterclockwise (Fig. 2a) and clockwise (Fig. 2b) vortex, respectively. Magnetization spatial distributions with $\varphi_0$ equal to 0 and $\pi$ can be transformed to each other by a spin coordinate system rotation around the z-axis by the angle $\pi$.

The magnetization distribution Eq. (10) has cylindrical symmetry, but does not have symmetry with respect to the reflection in the plane perpendicular to the cylinder base (the replacement $y \to -y$ without a change of local magnetization vectors direction). Under the reflection in the mentioned plane, the vortex magnetization distribution changes into an antivortex distribution. Note that the antivortex magnetization distribution cannot be obtained from the vortex by any spin coordinate rotation, and vice versa.

Note also that in the center of the disc, the magnetization vector has a component perpendicular to the cylinder base. This leads to the existence of a magnetic field outside of the particle. However, magnetic induction outside the cylinder has the same symmetry as the magnetization distribution, and can be presented in the form

$$\vec{B} = \begin{pmatrix} b(\rho,z)\sin(\varphi+\phi_0(z)) \\ -b(\rho,z)\cos(\varphi+\phi_0(z)) \\ b_z(\rho,z) \end{pmatrix}. \tag{11}$$

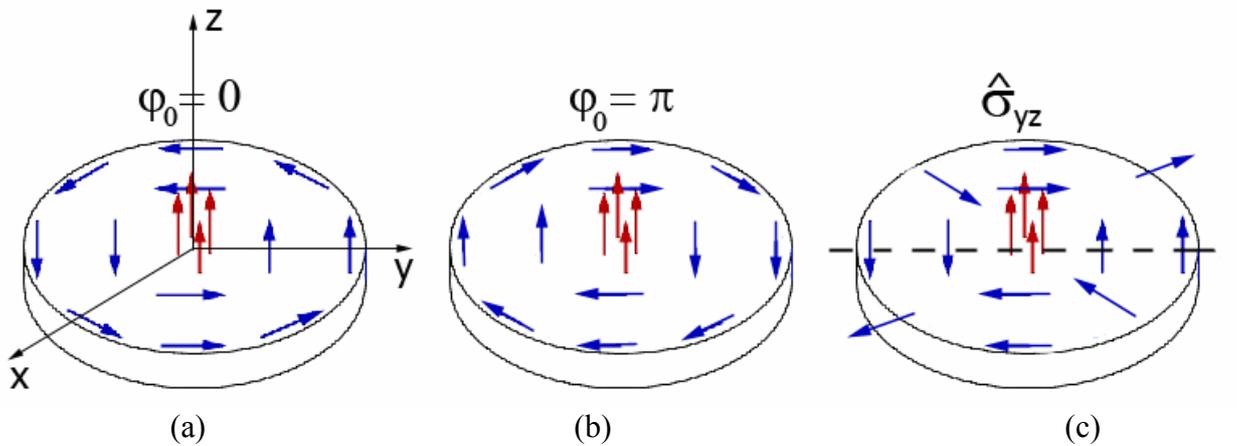

FIG. 2 (Color online). a), b) – particles with clockwise and counterclockwise vortex; c) – reflection of the vortex in the plane (yz) perpendicular to the cylinder base (antivortex state).

The expression for the unpolarized neutron elastic scattering cross section in the second order (first order does not lead to non-reciprocity and we omit it here) of perturbation theory has the form

$$\rho^{(3)}(\vec{k},\vec{k}') = -2i\left(\frac{m\mu_n}{2\pi\hbar^2}\right)^3 \int \left(\vec{B}(\vec{r}_1)\cdot[\vec{B}(\vec{r}_2)\times\vec{B}(\vec{r}_3)]\right)$$

$$\times \frac{\exp[i(\vec{k}'(\vec{r}_2-\vec{r}_1)+\vec{k}(\vec{r}_1-\vec{r}_3)+k|\vec{r}_2-\vec{r}_3|)]}{|\vec{r}_2-\vec{r}_3|}d\vec{r}_1 d\vec{r}_2 d\vec{r}_3 + c.c. \qquad (12)$$

Let us analyze the symmetry properties of $\rho^{(3)}(\vec{k},\vec{k}')$. This term in the cross section leads to non-reciprocity. It can be seen that the substitutions $\vec{k}\to-\vec{k}'$ and $\vec{k}'\to-\vec{k}$ are equal to the replacement

$$\vec{r}_2 \to \vec{r}_3,\ \vec{r}_3 \to \vec{r}_2, \qquad (13)$$

This in turn leads to the change of scalar triple product sign in the (12) resulting in the break of reciprocity

$$\rho^{(3)}(\vec{k},\vec{k}') = -\rho^{(3)}(\vec{k}',\vec{k}). \qquad (14)$$

Using the same logic it can be shown that detailed balance of vortex particle scattering is broken and $\sigma(\vec{k},\vec{k}',\vec{B}) \neq \sigma(\vec{k}',\vec{k},\vec{B})$. The latter leads to the asymmetry of the neutron scattering with respect to the reflection in the plane formed by the wave vector of the neutron and its projection on the cylinder base plane (incident plane). Let the neutron wave vector $\vec{k}$ lie in the plane (xz) (the coordinate system is the same as in fig. 2). We compare the cross section of two scattered neutrons, with wave vectors $\vec{k}_1$ and $\vec{k}_2$ arranged symmetrically with respect to the incident plane (fig. 3). By virtue of the cylindrical symmetry of the system, $\rho^{(3)}(\vec{k}_2,\vec{k}) = \rho^{(3)}(\vec{k},\vec{k}_1)$. Using Eq. (14), one can obtain $\rho^{(3)}(\vec{k},\vec{k}_2) = -\rho^{(3)}(\vec{k},\vec{k}_1)$. Thus, the scattering to the left and right from the incident plane is different (this is similar with skew scattering appearing for electrons in ferromagnets).

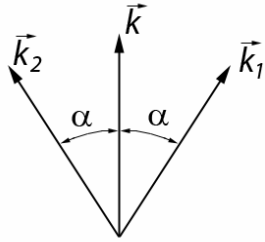

FIG. 3. Wave vectors of incident and scattered neutrons.

As follows from the property of the scalar triple product, $\rho^{(3)}(\vec{k},\vec{k}')$ does not depend on the $\varphi_0$ but depends on the magnetization direction in the center of the vortex.

The calculations of the scattering cross section for unpolarized neutrons interacting with magnetic vortex are done with the assumption concerning magnetization distribution inside the particle and magnetic induction outside the particle stated below. The magnetization has a non-zero uniform z-component in the small area near the cylinder axis. In the rest part of the disc, the magnetic moments lie in the plane (xy). Outside the particle, the magnetic induction has a z-component only in the region close to the cylinder axis, and the x- and y-components of magnetic induction are zero. The calculations are carried out for cobalt particles with different radii (20-100 nm) and with a width of 20 nm. The neutron wavelength is 1 nm, $k_z = -k'_z = 5.86$ nm$^{-1}$ (coordinate system is the same as Fig. 2), and $k_y = 0$, $k_x = k_\perp$, $k'_x = k_\perp \cos(\alpha)$, $k'_y = k_\perp \sin(\alpha)$. The particle radius is 70 nm. Fig. 4 presents the schematic representation of the cross section.

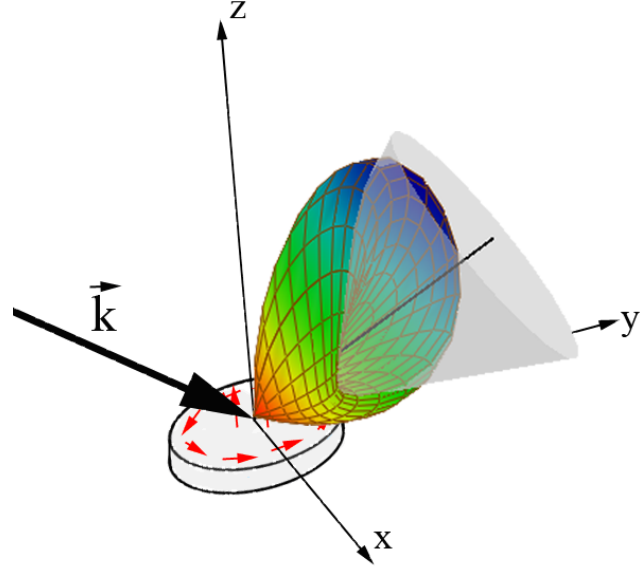

FIG. 4. Schematic representation of a neutron beam scattered by the vortex. $\vec{k}$ is incident neutron wave vector. It is situated in (yz) plane. Scattered neutrons mostly move to the left of the (yz)-plane.

Calculations show that in the considered regions of $\vec{k}$ and $\vec{k}'$ the relative magnitude of $\sigma(\vec{k},\vec{k}',\{\vec{B}(\vec{r})\}) - \sigma(-\vec{k}',-\vec{k},\{\vec{B}(\vec{r})\})$ is very small (about $10^{-5}$), and does not essentially depend on the particle sizes.

Note, however, that the symmetry considerations are correct not just in the field of application of the first Born approximation. For example, in the range of small $k_z$ for vortex particles, a practically total reflection should occur for neutrons. Although this situation lies beyond the Born approximation, detailed balance breaking and non-reciprocity should occur. The magnitude of the effects in the "total reflection" region can be essentially greater than in the region of big $k_z$, as in the situation with three mirrors considered in the next section. Also, the effects can be amplified by using the particle lattice, due to interference between the neutrons scattered by different particles.

The vortex magnetization distribution is realized in ferromagnetic particles [16] and in the crystals of MnSi under certain conditions [17]. In the first case, the particle diameter is about 100 nm and the width is about 30 nm. The vortex size in MnSi is about 20 nm.

Thus in this section we show that the non-reciprocity occurs in unpolarized neutron elastic scattering by a particle with a vortex magnetization distribution.

## IV. NON-RECIPROCITY OF THE NEUTRON REFLECTION BY THREE MIRRORS SYSTEM

In this section we consider another system with a non-coplanar magnetic field, namely, three magnetic mirrors. The magnetization vectors and configuration of the mirrors are shown in fig. 5. The surfaces of the mirrors are perpendicular to the xy-plane. If the magnetization vector of the central mirror is parallel to the z-axis, and the magnetization vectors of side mirrors are in the xy-plane, then the system has a non-coplanar field distribution. Consider a consecutive reflection from all three mirrors. The neutron reflection by a single mirror can be described by the following operator

$$\hat{M} = \begin{pmatrix} \dfrac{1-b_-}{1+b_-} & 0 \\ 0 & \dfrac{1-b_+}{1+b_+} \end{pmatrix}, \quad (15)$$

where $b_\pm = \sqrt{1 + 2m\dfrac{-V_{nuc} \pm \mu_n B}{k^2 \hbar^2 \sin^2 \varphi}}$, $m$ is the neutron mass, $\mu_n$ is the neutron magnetic moment, $k$ is the neutron wave number, $B$ is the mirror internal magnetic induction, $\varphi$ is the grazing angle of incidence, and $V_{nuc}$ is the spin-independent nuclear potential [11]. Operator $\hat{M}$ connects the amplitude of the incident and the reflected waves. In Eq. (15), it is assumed that the spin quantization axis is co-directional with the mirror magnetization vector.

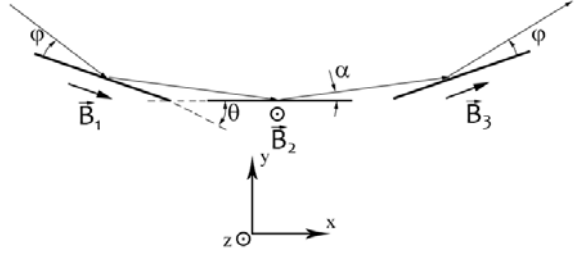

FIG. 5. Neutron beam reflection by the system of three magnetic mirrors.

Let us introduce the $\hat{O}_+$ operator describing the reflection from the whole system, in the case where the beam reflects from the mirrors in the order 1-2-3. This operator is expressed as follows

$$\hat{O}_+ = \hat{M}_3 \hat{R}_{32} \hat{M}_2 \hat{R}_{21} \hat{M}_1 \quad (16)$$

The grazing angles differ for each mirror, resulting in three different operators $\hat{M}_j$.

$$\hat{R}_{ij} = \exp(i(\hat{\vec{\sigma}} \cdot \vec{n}_{ij})\phi_{ij}/2) \quad (17)$$

is a spin coordinate system rotation operator. The rotation is accomplished by passing from the $i$-mirror to the $j$-mirror. Note that the quantization axis is co-directional with each mirror's magnetization. $\vec{n}_{ij}$ is the unit vector determining the rotation axis, and $\phi_{ij}$ is the rotation angle. It is straightforward to see that $\vec{n}_{ij} \perp (\vec{B}_i - \vec{B}_j)$. For example, $\vec{n}_{ij}$ can be written as follows

$$\vec{n}_{ij} = [\vec{B}_i \times \vec{B}_j]/(|\vec{B}_i| \cdot |\vec{B}_j|). \quad (18)$$

The turn operator has the following properties: $\hat{R}_{ij}^{-1} = \hat{R}_{ji} = \hat{R}_{ij}^+$ (index "+" refers to the Hermitian conjugation). It is significant that $\vec{n}_{ij}$ depends on the $j$-mirror coordinates. Therefore,

$\vec{n}_{32}$ depends on $\vec{n}_{21}$, i.e., it depends on the first rotation axis direction. We introduce vector $\vec{\tilde{n}}_{32}$ showing the direction of the second rotation $\hat{R}_{32}$ in the spin coordinate system connected with the first mirror

$$\vec{n}_{32} = \hat{\tilde{R}}_{21}\vec{\tilde{n}}_{32}, \tag{19}$$

where $\hat{\tilde{R}}_{21}$ is the rotation operator corresponding to the $\hat{R}_{21}$ rotation.

The intensity of the unpolarized neutron beam reflection can be found as half of the following matrix $I_+ = Tr[\hat{O}_+^+ \hat{O}_+]/2$. Its explicit form is

$$I_+ = Tr[\hat{M}_1^+ \hat{R}_{12} \hat{M}_2^+ \hat{R}_{23} \hat{M}_3^+ \hat{M}_3 \hat{R}_{32} \hat{M}_2 \hat{R}_{21} \hat{M}_1]/2 \tag{20}$$

Let us reverse the magnetic field direction. It is clear that the vectors $\vec{n}_{21}$ and $\vec{\tilde{n}}_{32}$ and the angles $\phi_{ij}$ are not changed under the reversal. This reversal is equivalent to the replacement of $b_+ \to b_-$ and $b_- \to b_+$ in Eq. (15). The latter can be described as the following transformation: $\hat{M} \to \sigma_x \hat{M} \sigma_x$. Using this transformation, it can be seen that the magnetic field reversal is also equivalent to the transformation $\hat{R}_{ij} \to \sigma_x \hat{R}_{ij} \sigma_x$ in Eq. (20). The operator $\hat{R}_{ij}$ commutates $\sigma_x$ only if the first rotation is around $x$-axis. This condition can always be met for the first rotation because the spin coordinates can be selected arbitrarily. Suppose that the first rotation axis is co-directional with the $x$-axis. Using Eq. (19), one can show that the vector $\vec{n}_{32}$ will be co-directional with the $x$-axis only if the vector $\vec{\tilde{n}}_{32}$ is also co-directional with the $x$-axis. The parallel alignment of $\vec{n}_{21}$ and $\vec{\tilde{n}}_{32}$ in combination with Eq. (18) indicates that all the mirror magnetization vectors are co-planar. Thus, the reflection coefficient is not changed if magnetization vector directions are reversed, in the case where they are co-planar. This is in agrement with the symmetry considerations mentioned in the Introduction. When the magnetizations are non-coplanar, the non-reciprocal reflection of the unpolarized neutron beam is possible.

Consider the situation where the beam reflection occurs in the opposite direction, i.e. it reflects in the order 3-2-1. In this case, the intensity of the reflection is

$$I_- = Tr[\hat{M}_3^+ \hat{R}_{32} \hat{M}_2^+ \hat{R}_{21} \hat{M}_1^+ \hat{M}_1 \hat{R}_{12} \hat{M}_2 \hat{R}_{23} \hat{M}_3]/2. \tag{21}$$

The trace is independent of the multiplication order, so Eq. (21) can be written as

$$I_- = Tr[\hat{M}_1 \hat{R}_{12} \hat{M}_2 \hat{R}_{23} \hat{M}_3 \hat{M}_3^+ \hat{R}_{32} \hat{M}_2^+ \hat{R}_{21} \hat{M}_1^+]/2. \tag{22}$$

Comparing Eq. (20) and Eq. (22), it is straightforward to see that the intensities of the normal and the reverse reflections are equivalent if the following equation holds true

$$\hat{M}_2^+ = \hat{M}_2 \tag{23}$$

The commutative property $[\hat{M}^+, \hat{M}] = 0$ is used. Eq. (23) is true only if the second mirror grazing angle is larger than the second critical angle

$$\varphi_{cr2} = \arcsin \frac{\sqrt{2m(\mu_n B + V_{nuc})}}{k\hbar}. \tag{24}$$

Thus, the reflection is not non-reciprocal when the second mirror grazing angle is larger than the second critical angle, even if the magnetizations are non-coplanar. Also, the reflection will be reciprocal if the following conditions are satisfied

$$\hat{M}_j \hat{M}_j^+ = 1 \text{ и } \hat{M}_i \hat{M}_i^+ = 1 \; (i \neq j). \tag{25}$$

These equations mean the grazing angles at which neutron beam hits to any of two mirrors are smaller than the first critical angle

$$\varphi_{cr1} = \arcsin \frac{\sqrt{2m(-\mu_n B + V_{nuc})}}{k\hbar}, \tag{26}$$

The situation corresponding to Eq. (25) is equivalent to the reflection of the system consisting of only two mirrors.

It is significant that the non-reciprocity mechanism in the three mirrors reflection differs from the mechanism considered in the third section of the paper. In case of scattering by magnetic vortex, the interference between the single and double scattered waves is important for the occurrence of non-reciprocity. There are no interference effects in the three mirrors system. It is essential here for the $\hat{M}$ operators to be **not** unitary (contrary to the situation of [18] where polarization spinners are considered). This leads to a change in the beam intensity that does not occur in the perturbation theory referred to in the second and third section of the paper. The single scattering leads only to the neutron spin rotation, and changes the phase of the wave function.

Non-reciprocity for the three mirrors system can be simply understood, if one considers the system to consist of two spin polarizer's and a spin rotator. The spin polarizer in terms of the nuclear and magnetic potentials means that $V_{nuc} = V_{mag} = \mu_n B$. Such a situation is realized, for example, in Fe-Co alloys. A Fe mirror can be used as a spin rotator. Imagine that the two side mirrors (polarizer's) are perpendicular, and that the middle mirror rotates neutrons spin by 90 degrees (Fig. 6). The rotation direction of neutrons spin by the second mirror does not depend on the direction of incidence. If the neutron beam falls from the left (Fig. 6a), then after reflection by the first two mirrors the beam will be polarized co-directionally with the third mirror magnetization. Therefore, the beam is fully reflected from the third mirror. If the beam falls from the right, then it will be counterdirected with respect to the third mirror magnetization, and it will be fully transmitted through the third mirror. Thus, the non-reciprocity in such a system is 100%.

In the physical sense, breaking of Eq. (23) indicates that central mirror rotates neutron spin polarization. From Eq. (15) shows that if Eq. (23) occurs, the middle mirror does not rotate the polarization vector of the neutron beam.

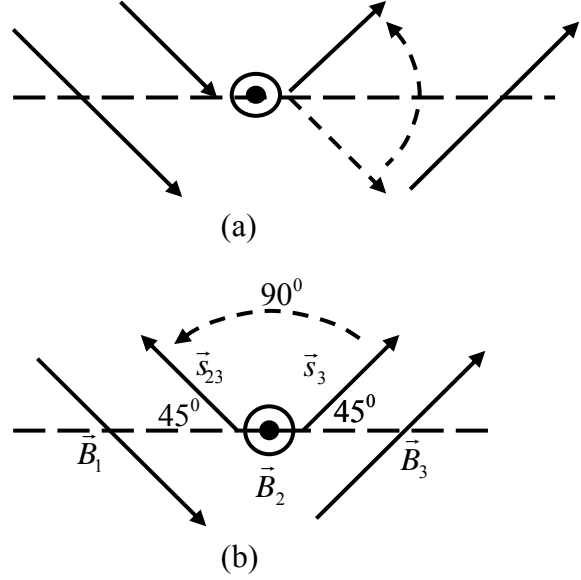

(a)

(b)

FIG. 6. Incidence from left (a) and right (b). $\vec{s}_1$ and $\vec{s}_3$ are polarization vectors after reflection by the first mirror. $\vec{s}_{21}$ and $\vec{s}_{23}$ are polarization vectors for each beam after reflection (spin rotation) by the second mirror.

In this work, we perform the calculation of the reflection coefficients of the unpolarized neutrons by system consisting of two side polarizer's and one central spin rotator. Two side polarizer's are placed symmetrically with respect to the central rotator. The angles between the mirrors are equal, and $\theta = 2°$. The polarizer's' grazing angles are equal to $\varphi = 1°$, and the rotator's grazing angle is $\alpha = \theta - \varphi$. The neutron wavelength varies in the range $\lambda = 10 \div 20$ Å. The directions of magnetization vectors are as shown in Fig. 5. The calculation results are presented in Fig. 7.

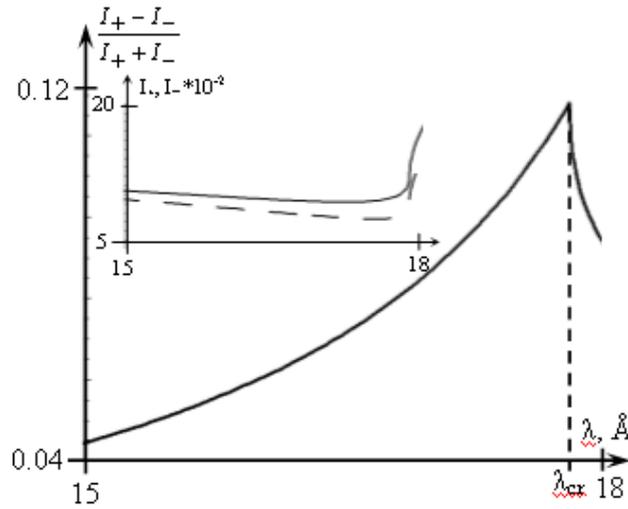

FIG. 7. The relative non-reciprocity of the system consisting of two side polarizers and one central spin rotator. The neutron beam reflectivity of the system consisting of two side polarizers and one central spin rotator. $\lambda_{cr}$ corresponds the critical angle (26) of the central spin rotator.

The wavelength range is selected in accordance with condition (23) so that the reflection is non-reciprocal in this range. The maximum of non-reciprocity is achieved at wavelength about 18 Å. The reflectivity at this wavelength is about 25%. The magnitude of non-reciprocity is about 3.5% of the incident beam intensity and about 12% of the reflected beam intensity. The magnitude of the effect is big enough to be observed experimentally [19]. For our knowledge experiments with two magnetic mirrors were performed earlier [20,21], therefore one can expect that three mirror system can also be realized.

At the end of this section, we note that this paper does not consider absorption. The absorption should also lead to non-reciprocity because it breaks the equality Eq. (23). Furthermore in the presence of the absorption, non-reciprocity should appear at all ranges of grazing angles, not only in the sub-critical range.

Thus, in this section, it is shown that the non-reciprocity appears in the unpolarized neutron beam reflection, in a system consisting of three magnetic mirrors whose magnetization vectors are non-coplanar.

## VI. CONCLUSION

The present work considers theoretically the elastic scattering of unpolarized neutrons by systems with non-coplanar magnetic induction spatial distributions. It is demonstrated that scattering by these systems is non-reciprocal. It is shown that this phenomenon occurs in the second order of perturbation theory.

Two real systems are described in which non-reciprocity occurs while neutrons are scattered. One of them is the three mirrors system, and the other is a ferromagnetic cylindrical particle with a vortex magnetization distribution. Calculations show that the non-reciprocity appearing in the three mirrors system is about 12% of the reflected beam intensity. Relative magnitude of non-reciprocity for scattering by a vortex particle is about $10^{-5}$.

This work was supported by "Dynasty" Foundation.

[1]L. D. Landau and E. M. Lifshitz, *Course on Theoretical Physics: Quantum Mechanics Non-Relativistic Theory*, (Pergamon, New York, 1977), Vol. 3.
[2]V.I. Belinicher, B.I. Sturman, Usp. Phys. Hauk, **130**, 3, 415 (1980).
[3]David S. Saxon, Phys. Rev., 100, 6, 1771 (1955)
[4]R.J. Potton, Rep. Progr. Phys., **67**, 717 (2004)
[5]A.L. Shelankov, G.E. Pikus, Phys. Rev. B, **46**, 6, 3326 (1992)
[6]P. Hillion, J. Opt. 9, 173 (1978)
[7]L. I. Mandel'shtamm, *Lectures on Optics, Relativity, and Quantum Mechanics*, (Nauka, Moscow, 1972), p. 9-10.
[8]L. Remer, B. Luthi H. Sauer, et. al., Phys. Rev. Lett., **56**, 25, 2752 (1986).
[9]L. Remer, E. Mohler W. Grill, et. al., Phys. Rev. B, **30**, 6, 3277 (1984).
[10]R.L. Stamps, B.L. Johnson R.E. Camley, Phys. Rev. B, **43**, 4, 3626 (1991).
[11]I. I. Gurevich, L. V. Tarasov, *Physics of Low Energy Neutrons*, (Nauka, Moscow, 1965).
[12]A.F. Andreev, V.I. Marchenko, Sov Phys Uspekhi, **23** (1), 21 (1980)
[13]L.D. Landau, E.M. Lifshitz, *Electrodynamics of Continuous Media* (v. 8 of *Course of Theoretical Physics*)(2ed., Pergamon, 1984)
[14]A.A. Fraerman, O.G. Udalov, JETP, **104**, 1, 62 (2006).
[15]V.K. Ignatovich, Yu. V. Nikitenko, A.A. Fraerman, JETP, **110**, 5, (2010)
[16]R.P. Cowburn, D.K. Koltsov, A.O. Adeyeye et al., Phys. Rev. Lett., **83**, 1042 (1999).
[17]S. Mühlbauer, B. Binz, F. Jonietz, et. al., Nature, **323**. 5916, 915 (2009).
[18]Y. Hasegawa, G. Badurek, Phys. Rev. A, **59**, 6, 4614 (1999).
[19]V.M. Pusenkov, N.K. Pleshanov, V.G. Syromyatnikov, et. al., JMMM, **175**, 237 (1997)
[20]S.G.E. te Velthuis, G.P. Feltcher, P. Blomquist, et. al., J. Phys.: Condens. Matter, 13, 5577 (2001)
[21]V.-O. de Haan, J. Plomp, T.M. Rekveld, et. al., Phys. Rev. Lett., **104**, 010401 (2010)